\documentclass[preprint,showpacs,preprintnumbers,amsmath,amssymb]{revtex4}
\usepackage{amssymb}

\usepackage{graphicx}
\usepackage{dcolumn}
\usepackage{bm}

\begin{document}

\title{Experimental characterization of entanglement dynamics in noisy channels}
 \author{Jin-Shi Xu}
 \affiliation{Key Laboratory of Quantum Information,
  University of Science and Technology
  of China, CAS, Hefei, 230026, People's Republic of China}
\author{Chuan-Feng Li$\footnote{email: cfli@ustc.edu.cn}$}
\affiliation{Key Laboratory of Quantum Information,
  University of Science and Technology
  of China, CAS, Hefei, 230026, People's Republic of China}
   \author{Xiao-Ye Xu}
\affiliation{Key Laboratory of Quantum Information, University of
Science and Technology of China, CAS, Hefei, 230026, People's
Republic of China}
\author{Cheng-Hao Shi}
\affiliation{Key Laboratory of Quantum Information, University of
Science and Technology of China, CAS, Hefei, 230026, People's
Republic of China}
\author{Xu-Bo Zou}
 \affiliation{Key Laboratory of Quantum Information,
  University of Science and Technology
  of China, CAS, Hefei, 230026, People's Republic of China}
 \author{Guang-Can Guo}
\affiliation{Key Laboratory
of Quantum Information, University of Science and Technology of
China, CAS, Hefei, 230026, People's Republic of China}
\date{\today }

\begin{abstract}
The dynamics of entanglement between two photons with one of them
passing through noisy quantum channels is characterized. It is
described by a simple factorization law which was first
theoretically proposed by Konrad {\it et al.} [Nature Phys., 4, 99
(2008)]. Quantum state tomography process is employed to reconstruct
the reduced density matrixes of the final states and the
corresponding concurrences are calculated. Good fittings between
experimental results and theoretical predictions are found, which
imply the validity of the general factorization law in the
characterization of entanglement dynamics.
\end{abstract}

\maketitle

Entanglement is the crucial resource for quantum communication and
computation \cite{Nielsen00}. However, due to unavoidable couplings
with the environment, it is readily destroyed when entangled
particles transmitted through noisy quantum channels \cite{Zurek03}.
Characterization the dynamics of entanglement becomes an important
task for those entanglement-based quantum information processing
protocols and it is also crucial for understanding the distinct
properties of entanglement dynamics, such as the phenomenon of
entanglement sudden death \cite{Yu04,Yu06,Almeida07,Yu09,Xu08_1}.

In order to quantify the entanglement, we need a convenient quantity
to measure it. The Wootter's concurrence \cite{Wootters98} is a good
choice for two qubits, which is given by
\begin{equation}
C=\textrm{max}\{0,\Gamma\}, \label{eq:concurrence}
\end{equation}
where
$\Gamma=\sqrt{\lambda_{1}}-\sqrt{\lambda_{2}}-\sqrt{\lambda_{3}}-\sqrt{\lambda_{4}}$
and the $\lambda_{i}$ are the eigenvalues in decreasing order of the
matrix
$\rho(\sigma_{y}\otimes\sigma_{y})\rho^{*}(\sigma_{y}\otimes\sigma_{y})$.
$\sigma_{y}$ is the second Pauli matrix and $\rho^{*}$ is the
complex conjugate of $\rho$ in the canonical basis $\{|00\rangle,
|01\rangle, |10\rangle, |11\rangle\}$. The concurrence ranges from 0
(the separated state) to 1 (the maximally entangled state).

Usually, the tomography process \cite{James01} which requires
measurement on a complete set of observable quantities is used to
reconstruct the final density matrix $\rho$ and the entanglement of
the state is always nonlinear dependent on it \cite{Mintert05}, just
as we have shown above. Therefore, the knowledge of the dynamics of
entanglement can only deduce from the time evolution of the state
\cite{Yu04,Yu06,Almeida07,Xu08_1,Zyczkowski01,Carvalho04,Dur04,Carvalho07}.
The general result on entanglement dynamics is still a great
challenge.

Recently, the work of Konrad \textit{et al.} gives a general
factorization law on the entanglement dynamics of the biparticle
system under the action of an arbitrary channel on one of the
component \cite{Konrad08}. The main result they got is the
formulation:
\begin{equation}
C[(\openone\otimes\$)|\chi\rangle\langle\chi|]=C[(\openone\otimes\$)|\phi^+\rangle\langle\phi^+|]C(|\chi\rangle),
\label{pure}
\end{equation}
where $|\chi\rangle$ is the initial input pure state and
$|\phi^+\rangle$ is the maximally entangled state. We use $C_{LP}$
to represent the left term of the equality and $C_{RP}$ to represent
the right term, for simplicity. It is clearly shown that the
entanglement dynamics in the one-sided noisy channel only relate to
the dynamics of the maximally entangled state in the channel with a
factor of the concurrence of the initial state $|\chi\rangle$.

Due to the convexity property of concurrence \cite{Wootters98}, this
factorization law can be generalized to the mixed initial state
$\rho_{0}$ \cite{Konrad08}
\begin{equation}
C[(\openone\otimes\$)\rho_{0}]\leq
C[(\openone\otimes\$)|\phi^+\rangle\langle\phi^+|]C(\rho_{0}).
\label{onemixed}
\end{equation}
We also use $C_{LM}$ and $C_{RM}$ to represent the left and the
right terms of this inequality, respectively. This inequality gives
the upper bound of the evolved entanglement.

Here, using an all optical setup, we experimentally characterize the
two-photon entanglement dynamics with one of the photons passing
through the phase damping channel or the amplitude decay channel.
The experimental results show that the time evolution of
entanglement is fully characterized by the channel's action on the
maximally entangled state and we verify the factorization laws.

\begin{figure}[tbph]
\begin{center}
\includegraphics [width= 3.0in]{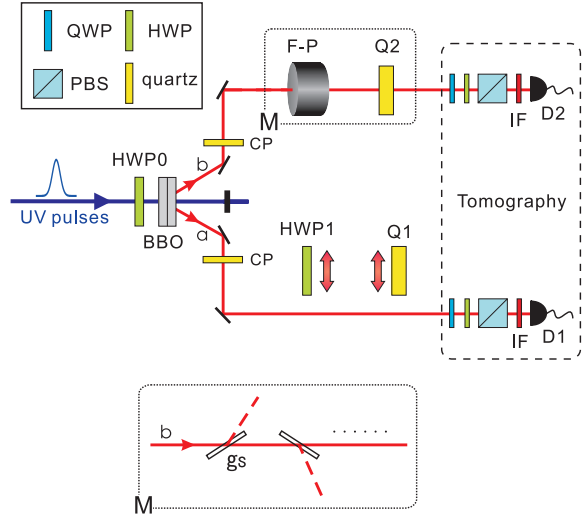}
\end{center}
\caption{(Color online). The experimental setup to investigate the
entanglement dynamics in different noisy channels. Entangled
polarization photon pairs generated from the adjacent nonlinear
crystals (BBO) emit into modes $a$ and $b$. The half-wave plate
(HWP0) is used to change the polarization of the pump light. A
half-wave plate (HWP1) and quartz plates (Q1) are inserted into mode
$a$ depending on different cases. The dotted pane M represents the
noisy channels. The Fabry-Perot (F-P) cavity followed by quartz
plates (Q2) simulates the phase damping channel, while the sets of
glass slabs (gs) which are tilted to the Brewster angles represent
the amplitude decay channels (the dashed lines represent the
reflected photons). Quarter-wave plates (QWP), half-wave plates
(HWP) and polarization beam splitters (PBS) in both arms are used to
set the detecting polarization bases for the state reconstruction.
Both photons are finally detected by silicon avalanche photodiodes
(D1 and D2) equipped with 3 nm interference filters to give
coincident counts.} \label{fig:setup}
\end{figure}

Our experimental setup is shown in Figure 1. Ultraviolet (UV) pulses
with wavelength centered at 390 nm which are frequency doubled from
a Ti:sappire laser with a pulse width of 130 fs and a repetition
rate of 76 MHz pump into a two geometry type I beta-barium-borate
(BBO) crystals to generate entangled two photons \cite{Kwiat99}. The
polarization of these UV pulses is set by the half-wave plate HWP0
which can be used to prepared different pure input states
$|\chi\rangle=\alpha|HH\rangle_{a,b}+\beta|VV\rangle_{a,b}$
($\alpha$ and $\beta$ are the relative amplitudes which are set to
be real for simplicity and they obey the relationship of
$\alpha^{2}+\beta^{2}=1$). Quartz plates (CP) in both emitted modes
$a$ and $b$ are used to compensate the temporal difference between
horizontal and vertical polarization components in these two
nonlinear crystals \cite{Xu06}.

A Fabry-Perot (F-P) cavity followed by quartz plates Q2 in the
dotted pane M is used to simulate the phase damping channel, which
has been used in our previous experiment \cite{Xu08_1}. When the
photon passes through Q2 with thickness $L$, the relative phase
between $H$- and $V$- polarization photon can be calculated as
$\phi=L\Delta n\omega/c$, where $c$ is the velocity of the photon in
the vacuum. $\Delta n=n_{o}-n_{e}$ is the difference between the
indexes of refraction of ordinary light ($n_{o}$) and extraordinary
light ($n_{e}$), which can be treated as a constant of 0.01 for the
small frequency distribution. The amplitude decay channel shown also
in the dotted pane M is simulated by a sets of glass slabs (gs)
which are tilted to the Brewster angles (about $57^{\circ}$)
\cite{Kwiat01}. After passing through the slaps, the photon with
vertical polarization has some probability to reflect and the photon
with horizontal polarization will transmit completely. Therefore,
the corresponding quantum map \cite{Preskill98} can be written as
\begin{eqnarray}
&|H\rangle_{S}|0\rangle_{E}\longrightarrow|H\rangle_{S}|0\rangle_{E},&
\notag\\
&|V\rangle_{S}|0\rangle_{E}\longrightarrow\sqrt{1-\epsilon}|V\rangle_{S}|0\rangle_{E}
+\sqrt{\epsilon}|H\rangle_{S}|1\rangle_{E}.&
\end{eqnarray}
where $|0\rangle_{E}$ and $|1\rangle_{E}$ represent the propagation
paths of the photon and $\epsilon$ represents the total reflectivity
of the glass slaps. We can change the polarization of the reflected
photon into $|H\rangle$ with half-wave plates and detect them with
single photon detectors (not shown in fig. \ref{fig:setup}).

A half-wave plate (HWP1) and quartz plates (Q1) are inserted into
mode $a$ depending on different cases to introduce different
decoherence effect on the maximally entangled state
$|\phi^{+}\rangle=\frac{1}{\sqrt{2}}(|HH\rangle_{a,b}+|VV\rangle_{a,b})$
to prepare different mixed input states. The optical axis of HWP1 is
set to be $22.5^{\circ}$ operating as a Hadamard gate and Q1 with
the optical axis set to be horizontal can induce relative phase in
the basis $H/V$.

The density matrixes of the final states are reconstructed by the
tomography process \cite{James01}. Experimentally, quarter-wave
plates (QWP), half-wave plates (HWP) and polarization beam splitters
(PBS) in both arms are used to set the standard 16 polarization
analysis measurement bases \cite{James01}. Both photons are detected
by silicon avalanche photodiodes D1 and D2 to give coincident
counts. Narrow band interference filters (IF) with a full width at
half maximum (FWHM) of 3 nm in front of detectors are used to reduce
the background and define the bandwidth of the photons.

\begin{figure}[tbph]
\begin{center}
\includegraphics [width= 3.0in]{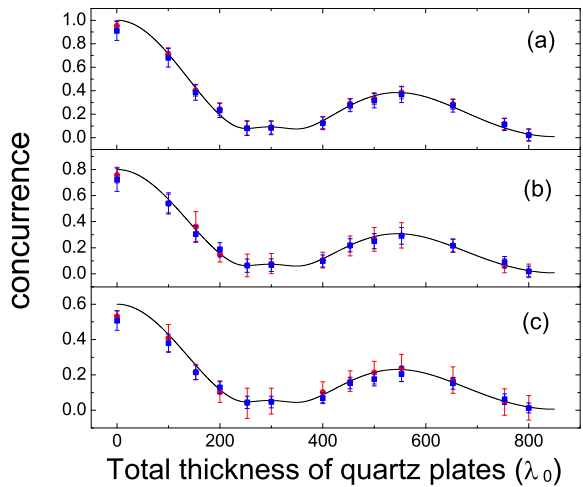}
\end{center}
\caption{(Color online). Experimental results for the entanglement
dynamics of different pure initial states in the phase-damping
channel. (a)$\alpha^{2}=\frac{1}{2}$. (b) $\alpha^{2}=\frac{1}{5}$.
(c) $\alpha^{2}=\frac{1}{10}$. Red dots and blue squares represent
$C_{LP}$ and $C_{RP}$, respectively. The solid lines and dashed
lines representing the theoretical predictions of $C_{LP}$ and
$C_{RP}$ completely overlap. $\lambda_{0}$=780 nm.}
\label{fig:phase-purechannel}
\end{figure}

\begin{figure}[tbph]
\begin{center}
\includegraphics [width= 3.0in]{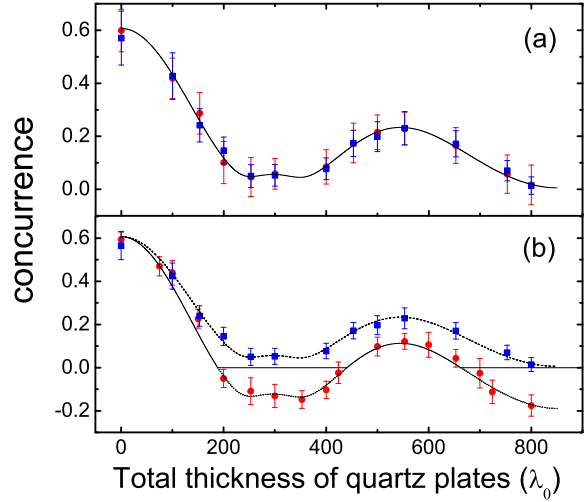}
\end{center}
\caption{(Color online). Experimental results for different mixed
input states. (a) the case with inserting only Q1=117$\lambda_{0}$
into mode $a$. (b) the case with inserting both the HWP1 and
Q1=117$\lambda_{0}$ into mode $a$. Red dots is the experimental
results of $\Gamma_{LM}$ and blue squares represent $C_{RM}$. The
solid lines and dashed lines are the theoretical predictions of
$C_{LM}$ and $C_{RM}$, respectively (they completely overlap in the
case of (a)). At the area of $\Gamma_{LM}\leq0$ where the
theoretical prediction is represented by dotted lines (b), the
concurrence is set to 0 according to equation (1).}
\label{fig:phase-mix}
\end{figure}

We first demonstrate the entanglement dynamics of different pure
input states in the phase-damping channel, as shown in fig. 2. Three
different pure input states $|\chi\rangle$ with
$\alpha^{2}=\frac{1}{2}$ (fig. 2(a)), $\alpha^{2}=\frac{1}{5}$ (fig.
2(b)), $\alpha^{2}=\frac{1}{10}$ (fig. 2(c)) are considered,
respectively. The x axis represents total thickness of Q2 which is
given by its retardation and $\lambda_{0}=780$ nm. Experimental
results of $C_{LP}$ and $C_{RP}$ (the product of
$C[(\openone\otimes\$)|\phi^+\rangle\langle\phi^+|]$ and
$C(|\chi\rangle)$) are represented by red dots and blue squares,
respectively. We can find that both of them overlap with each other
very well, which infers that $C_{LP}=C_{RP}$. The theoretical
predictions $C_{LP}$ (solid lines) and $C_{RP}$ (dashed lines) are
calculated from equation (\ref{eq:concurrence}) with the fitting
parameters the same as the case we demonstrate the phenomenon of
entanglement sudden death \cite{Xu08_1} and they completely overlap
with each other. It is also seen from fig. 2 that experimental
results agree well with theoretical predictions. In our experiment,
the pumping power is about 200 mW and the integration time is 6
minutes. Error bars are mainly due to the counting statistics and
the uncertainties in aligning the wave plates \cite{James01}.

When it cames to the case with mixed input states, the dynamics of
the maximally entangled state in the channel gives the upper bound
of the entanglement dynamics according to inequality
(\ref{onemixed}). Fig. \ref{fig:phase-mix}(a) shows the case with
the mixed input state prepared by only inserting Q1=$117\lambda_{0}$
into mode $a$ to dephase the maximally entangled state
$|\phi^{+}\rangle$ ($\alpha^{2}=\frac{1}{2}$). We can see that the
experimental results agree well with the theoretical prediction and
$C_{LM}=C_{RM}$. We further demonstrate the other case in which we
insert both the HWP1 and Q1=117$\lambda_{0}$. Fig.
\ref{fig:phase-mix}(b) shows our experimental results, where the
phenomenon of entanglement sudden death ($C_{LM}$) occurs
\cite{Xu08_1}. At the area of $\Gamma_{LM}\leq0$, the concurrence is
set to 0 corresponding to equation (1). It can be seen that $C_{LM}$
(red dots) is less than $C_{RM}$ (blue squares) during the
evolution, which is consistent with the theoretical predictions of
$C_{LM}$ (solid line) and $C_{RM}$ (dashed line).

\begin{figure}[tbph]
\begin{center}
\includegraphics [width= 3.0in]{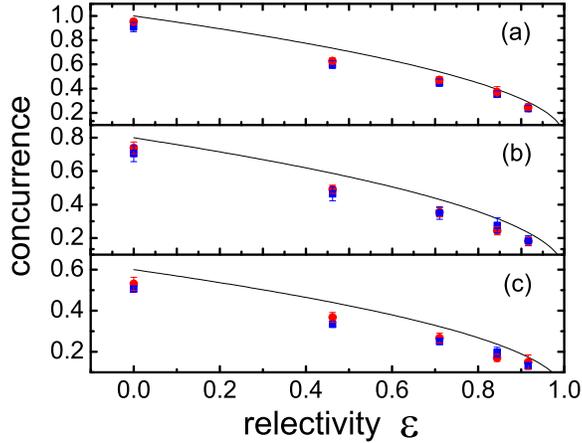}
\end{center}
\caption{(Color online). Entanglement dynamics of different pure
states in the amplitude decay channel. (a) $\alpha^{2}=\frac{1}{2}$.
(b) $\alpha^{2}=\frac{1}{5}$. (c) $\alpha^{2}=\frac{1}{10}$. Red
dots and blue squares represent experimental results of $C_{LP}$ and
$C_{RP}$, respectively. The corresponding theoretical predictions of
$C_{LP}$ (solid lines) and $C_{RP}$ (dashed lines) completely
overlap.} \label{fig:amplitude-pure}
\end{figure}

Now, we consider the entanglement dynamics in the amplitude decay
channel, where the F-P cavity and quartz plates Q2 in the dotted
pane M are replaced by the set of glass slaps titled to the Brewster
angles. By changing the number of the glass slaps, we can control
the reflectivity of this channel. In order not to disturb the
transmitted photons, two relatively placed slaps are added or
removed at the same time. The reflectivity of each two slaps we
measured is about 0.46. The counts of reflected photons is deduced
from the difference between the transmitted vertical polarization
photons without glass slaps and with glass slaps. The coincident
counts of the 16 measurement bases to reconstruct the final
photon-state are consisted by the corresponding transmitted and
reflected parts. The transmitted counts of each measurement basis
are detected by the tomography measurement setup in fig.
\ref{fig:setup}, while the corresponding reflected counts is
calculated as the product of the total counts of the reflected
photons and the theoretical detecting probability of each
measurement basis in the reflected photon-state.

Fig. \ref{fig:amplitude-pure} shows the case of the amplitude decay
channel with different pure input states $|\chi\rangle$. In this
experiment, the reflected photon-state is $|VH\rangle_{a,b}$. Fig.
\ref{fig:amplitude-pure}(a), (b) and (c) correspond to the initial
state with $\alpha^{2}=\frac{1}{2}$, $\frac{1}{5}$ and
$\frac{1}{10}$, respectively. We can see that the experimental
results of $C_{LP}$ (red dots) are equal to $C_{RP}$ (blue squares).
The theoretical predictions of $C_{LP}$ (solid lines) and $C_{RP}$
(dashed lines) are both given by $2\alpha\beta\sqrt{1-\epsilon}$.
The difference between the experimental results and the
corresponding theoretical predictions comes mainly from disturbance
of the preparation of the initial state and the absorbtion of the
slabs.

\begin{figure}[tbph]
\begin{center}
\includegraphics [width= 3.0in]{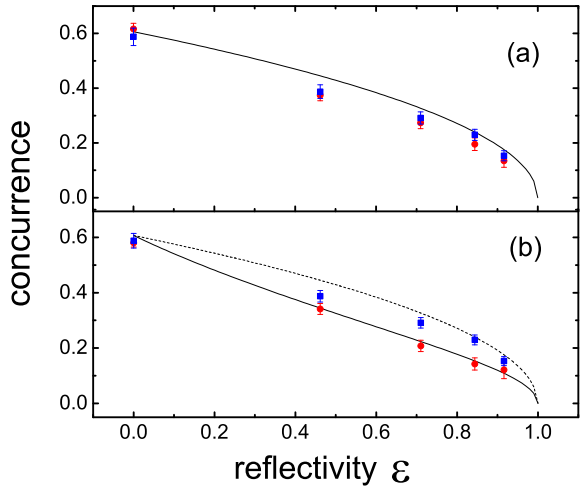}
\end{center}
\caption{(Color online). Entanglement dynamics of the amplitude
decay channel for mixed input states. (a) the case with inserting
only Q1=117$\lambda_{0}$ into mode $a$. (b) the case with inserting
the HWP1 and Q1=117$\lambda_{0}$ into mode $a$ at the same time.
Experimental results are represented by red dots ($C_{LM}$) and blue
squares ($C_{RM}$). Solid lines and dashed lines are the
corresponding theoretical fittings of $C_{LM}$ and $C_{RM}$ (they
completely overlap in the case of (a)).} \label{fig:amplitude-mix}
\end{figure}

Fig. \ref{fig:amplitude-mix} shows the experimental results of the
amplitude decay channel for mixed input states. The initial mixed
state in fig. \ref{fig:amplitude-mix}(a) is the same as that in the
case of fig. 3(a), which is prepared by inserting only
Q1=117$\lambda_{0}$ into mode $a$ with $\alpha^{2}=\frac{1}{2}$. We
find that $C_{LM}$ (red dots) equal to $C_{RM}$ (blue squares). The
theoretical predictions of $C_{LM}$ (solid lines) and $C_{RM}$
(dashed lines) completely overlap and are consistent with the
experimental results. Therefore, the entanglement dynamics of the
inequality (\ref{onemixed}) reaches its upper bound. In another case
where the initial mixed state is prepared by inserting the HWP1 and
Q1=117$\lambda_{0}$ at the same time, we find that $C_{LM}\leq
C_{RM}$ and inequality (3) holds, as shown in fig.
\ref{fig:amplitude-mix}(b). In this experiment, the reflected
photon-state is given by $\rho_{r}=1/2(|HH\rangle\langle
HH|-\kappa_{a}^{\ast}|HH\rangle\langle
VH|-\kappa_{a}|VH\rangle\langle HH|+|VH\rangle\langle VH|)$, where
$\kappa_{a}$ is the decoherence parameter in mode $a$. Experimental
results agree with the corresponding theoretical prediction
employing equation (\ref{eq:concurrence}).

In conclusion, we have characterized the entanglement dynamics in
quantum channels with the maximally entangled state and verified the
entanglement factorization law \cite{Konrad08}. The results provide
a novel method to describe entanglement dynamics in noisy channels
and would have great importance on the construction of the complex
quantum network \cite{Kimble08}.

\textit{Note:} After we finished this manuscript, we find that
similar experimental results have been published online by
Far\'{\i}as \textit{et al.} \cite{Farias09}. They have characterized
the entanglement dynamics in the amplitude decay channel, which is
simulated by a modified Sagnac interferometer. In our experiment, we
demonstrate the characterization both in the phase damping channel
and the amplitude decay channel, which are simulated by the F-P
cavity followed by quartz plates and the glass slaps with the axes
titled to the Brewster angles, respectively. What is more, the
concurrence of the evolved maximally entangled state
($C[(\openone\otimes\$)|\phi^+\rangle\langle\phi^+|]$) is calculated
from the experimental form of the single channel $\$$ in their paper
\cite{Farias09}, while we measure all the quantities in equality
(\ref{pure}) and inequality (\ref{onemixed}), and then compare the
left terms and the right terms to verify the idea of characterizing
the entanglement dynamics in noisy channels with the maximally
entangled state.

We thank Y. X. Gong for his notification. This work was supported by
National Fundamental Research Program, the Innovation funds from
Chinese Academy of Sciences, National Natural Science Foundation of
China (Grant No.60121503, 10874162) and Chinese Academy of Sciences
International Partnership Project.

\end{document}